\documentclass[aps,reprint,amsmath,amssymb, floatfix]{revtex4-2}

\draft

\usepackage[T1]{fontenc}
\usepackage{mathptmx} 
\usepackage[english]{babel}
\usepackage[utf8]{inputenc}
\usepackage{polski}
\usepackage{indentfirst}
\usepackage{amsthm}
\usepackage{amsmath}
\usepackage{amssymb}
\DeclareMathOperator{\sinc}{sinc}
\usepackage{lmodern} 
\usepackage{graphicx}
\usepackage{geometry}
\usepackage[caption=false]{subfig}
\usepackage{hyperref}
\usepackage{relsize}
\usepackage[capitalize]{cleveref}
\usepackage{multirow}
\usepackage{bm,array}
\newgeometry{tmargin=1.5cm, bmargin=1.5cm, lmargin=1.5cm, rmargin=1.5cm}
\usepackage{xfrac}
\usepackage{graphicx}
\usepackage{tabularx}
\usepackage{booktabs}
\usepackage{float}
\usepackage{dcolumn}
\usepackage{etoolbox}

\begin{document}

\title{Spin-wave spectral analysis in crescent-shaped ferromagnetic nanorods}
\author{Mateusz Gołębiewski}\email{mateusz.golebiewski@amu.edu.pl}
\affiliation{Institute of Spintronics and Quantum Information, Faculty of Physics, Adam Mickiewicz University, Uniwersytetu Poznańskiego 2, 61-614 Poznań, Poland}
\author{Hanna Reshetniak}
\affiliation{Institute of Spintronics and Quantum Information, Faculty of Physics, Adam Mickiewicz University, Uniwersytetu Poznańskiego 2, 61-614 Poznań, Poland}
\author{Uladzislau Makartsou}
\affiliation{Institute of Spintronics and Quantum Information, Faculty of Physics, Adam Mickiewicz University, Uniwersytetu Poznańskiego 2, 61-614 Poznań, Poland}
\author{Arjen van den Berg}
\affiliation{School of Physics and Astronomy, Cardiff University, The Parade, Cardiff CF24 3AA, U.K.}
\author{Sam Ladak}
\affiliation{School of Physics and Astronomy, Cardiff University, The Parade, Cardiff CF24 3AA, U.K.}
\author{Anjan Barman}
\affiliation{Department of Condensed Matter and Materials Physics, S. N. Bose National Centre for Basic Sciences, Block JD, Sector III, Salt Lake, Kolkata-700 106, India}
\author{Maciej Krawczyk}
\affiliation{Institute of Spintronics and Quantum Information, Faculty of Physics, Adam Mickiewicz University, Uniwersytetu Poznańskiego 2, 61-614 Poznań, Poland}

\date{\today}

\begin{abstract}
The research on the properties of spin waves (SWs) in three-dimensional nanosystems is an innovative idea in the field of magnonics. Mastering and understanding the nature of magnetization dynamics and binding of SWs at surfaces, edges, and in-volume parts of three-dimensional magnetic systems enables the discovery of new phenomena and suggests new possibilities for their use in magnonic and spintronic devices. In this work, we use numerical methods to study the effect of geometry and external magnetic field manipulations on the localization and dynamics of SWs in crescent-shaped (CS) waveguides. It is shown that changing the magnetic field direction in these waveguides breaks the symmetry and affects the localization of eigenmodes with respect to the static demagnetizing field. This in turn has a direct effect on their frequency. Furthermore, CS structures were found to be characterized by significant saturation at certain field orientations, resulting in a cylindrical magnetization distribution. Thus, we present chirality-based nonreciprocal dispersion relations for high-frequency SWs, which can be controlled by the field direction (shape symmetry) and its amplitude (saturation).
\end{abstract}

\maketitle

\section{Introduction}
\label{Sec:int}

Today, the topic of spin waves (SWs) and their control in magnetic materials covers a broad spectrum of research. The technological potential of signal transport without the emission of Joule-Lenz heat~\cite{Serga2010YIGMagnonics, Yan2012MagnonicWires, Barker2016ThermalGarnet}, the wavelength of SWs from micrometers to tens of nanometers for frequencies from few GHz to several hundred GHz~\cite{Schneider2008RealizationGates, Kajiwara2010TransmissionInsulator, Yu2014MagneticNanomagnonics, Maendl2017SpinScale}, the ability to control the dispersion and group velocity of magnons~\cite{Garcia-Sanchez2015NarrowWalls, Wagner2016MagneticNanochannels, Duerr2012EnhancedWaveguide, Lan2015Spin-WaveDiode} and high energy efficiency without compromising the conversion speed~\cite{Chumak2015MagnonSpintronics, VKruglyak2010Magnonics, Serga2010YIGMagnonics, Lenk2011TheMagnonics}, make them a desirable successor to conventional electric currents, among others in computing, memory, and various types of microwave systems~\cite{Pirro2021AdvancesMagnonics, Chumak2019MagnonComputing, Nikitov2015Magnonics:Electronics, Mahmoud2020IntroductionComputing, Barman2021TheRoadmap, Chumak2022AdvancesComputing}.

It is promising to design advanced ma\-gno\-nic systems where, thanks to static (e.g., geometry, topology, material properties, magnetization texture) and dynamic factors (e.g., frequency of SWs, dynamic couplings, and direction of the external magnetic field), it is possible to control magnons and adjust their dynamics to the given goals. 
In ferromagnetic materials, the properties of SWs are determined by strong isotropic exchange interactions coexisting with anisotropic magnetostatic interactions. The localization of SWs is a natural consequence of the development and miniaturization of nanoscale magnetic systems and attempts to manipulate their magnetization. One example of such localizations is the edge mode~\cite{Jorzick2002SpinElements, Park2002SpatiallyWires, Bayer2003Spin-waveElements, Bailleul2003MicromagneticStripe, Bayer2004SpinStripe, McMichael2006EdgeEdges, Maranville2006CharacterizationDynamics, Kruglyak2006Timeinvited, Demidov2010NonlinearEllipses, Maranville2007VariationStripes, Zhu2010ModificationEdges, Shaw2009SpinDamping, Nembach2011EffectsNanomagnets, Kruglyak2005PicosecondPrecession, McMichael2008ThicknessStripes}, where SWs are bound or only propagate along the outer parts of a system. The strong heterogeneity of the internal demagnetizing field at the edges perpendicular to the magnetization allows the localization of SWs in these regions, and the localization allows the trapped wave modes to act as information carriers or sensitive probes of the magnetic properties of an entire system.

In recent years, there has been significant development of new fabrication techniques, such as two-photon lithography and focused electron beam induced deposition, which now allow the fabrication and analysis of complex 3D structures at the nanometer scale~\cite{Fernandez-Pacheco2017Three-dimensionalNanomagnetism, Donnelly2021ComplexNanostructures, Hunt2020HarnessingNanoscale, vandenBerg2022CombiningNanostructures, Fischer2020LaunchingNanostructures, Makarov2022NewNanoarchitectures}. Understanding the influence of geometric and topological properties on the propagation of SWs in 3D systems is at a very early stage of research. Exciting effects are shown by crescent-shaped (CS) nanowires arranged in diamond bond-like networks, i.e., enabling the analysis of states close to degeneration and providing a platform for reconfigurable magnonic devices~\cite{May2019RealisationLattice, Gartside2020Current-controlledCrystals, Sahoo2021ObservationStructure, May2021MagneticSpin-ice, Sahoo2018UltrafastStructure}. 
In the above research, the nanorods are building blocks of more complex systems, demonstrating the existing implementation of CS structures in experimental and theoretical studies. A single CS waveguide may also be of interest in its own right. Moreover, the knowledge of single nanowires will provide a better understanding of the complex dynamics of magnetization in 3D structures. entire system.

This study investigates the shape and curvature of CS nanowires to determine how they affect the magnetization dynamics. In this context, there are important studies showing effective Dzyaloshinskii-Moriya and anisotropy interactions associated with curvature, as proposed in Ref.~\cite{Streubel2016MagnetismGeometries}. There are also numerous other intriguing effects in curved magnetic wires and films~\cite{Sheka2021AMagnetism}, the understanding of which opens new research avenues and motivates the systematic analysis of the CS structures performed in this work. 

The geometry of the simulated structure is shown in Fig.~\ref{Fig:geom}, where the coordinate system and field direction are defined. The SW modes have different spatial distributions in the described structure, ranging from localized to volumetric, spreading throughout the volume, and having different quantization numbers and properties. 
We study two types of nanorods with CS cross sections -- those with rounded edges and those with sharp edges. With this approach, we combine the practical and theoretical analysis of eigenmodes and determine the range where and to what extent the contribution of the edge changes the results for the whole nanorod. In addition, the SW propagation and its dependence on the transverse localization of the modes have been analyzed. The obtained dispersion relations show interesting nonreciprocal properties that can be used for dynamic manipulation of SWs in these waveguides.

The structure of the paper is as follows. First, we describe the system geometry (Sec.~\ref{Sec:Geometry}) and the numerical methods used in the simulations (Sec.~\ref{Sec:mic_sim}). In Sec.~\ref{Sec:Results} we present the results and analyze the SW spectra. The results obtained in magnetization saturation at different external field orientations with respect to the CS nanorod axis are presented in Sec.~\ref{Sec:Angle_dependence}. The SW spectrum in the continuous transformation from elliptic to CS nanorod is shown in Sec.~\ref{Sec:Ellipse}, and we conclude how the edge sharpness influences the magnonic response of the structure in Sec.~\ref{Sec:Edge}. 
In Sec.~\ref{Sec:Unsaturated} we study the effect of decreasing magnetic field and static magnetization distribution. Finally, in Sec.~\ref{Sec:Dispersion} we examine the influence of the magnetic field orientation on the dispersion relation of the SWs along the long axis. The last section is a summary of the paper.

\section{Geometry and material parameters \label{Sec:Geometry}}

\begin{figure}[htp]
\includegraphics[width=\linewidth]{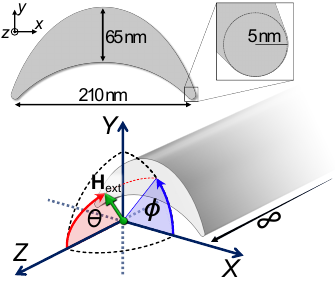}
\caption{Model of the nanorod with a CS section. Spherical coordinates for the external magnetic field $\textbf{H}_\text{ext}$ and the main dimensions are marked.
\label{Fig:geom}}
\end{figure}

In this research, infinitely long ferromagnetic nanorods with CS cross sections (Fig.~\ref{Fig:geom}) are studied. Thus, the magnetic properties (magnetization, demagnetizing field, etc.) are considered to be homogeneous along the $z$-axis. Such nanowires can be realized using a combination of two-photon lithography and evaporation~\cite{May2019RealisationLattice, Gartside2020Current-controlledCrystals, Sahoo2021ObservationStructure, May2021MagneticSpin-ice}.

\begin{figure*}[t]
\includegraphics[width=\linewidth]{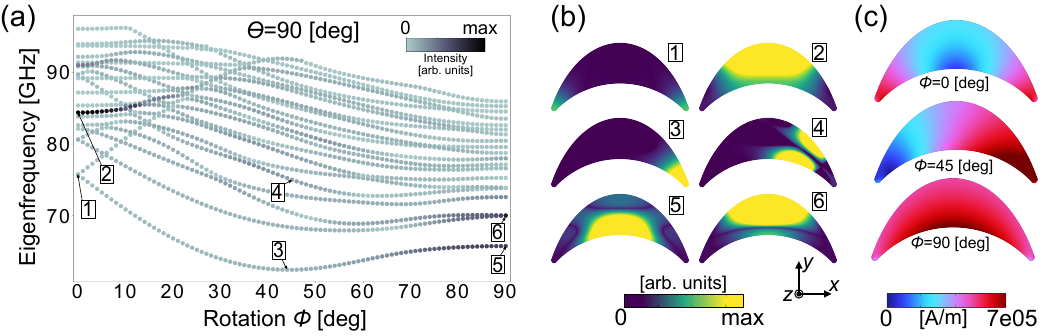}
\caption{Eigenanalysis of the CS nanowire as a function of the changing direction of the external magnetic field that saturates the magnetization, $\mu_0{H}_\text{ext}=3$~T. Plot (a) shows the dependence of the eigenfrequencies on the azimuthal angle $\phi$ in the range 0-90$^\circ$ for the polar angle $\theta=90^\circ$. The colors correspond to the intensities of the individual eigenmodes according to Eq.~(\ref{Eq:int}). In (b) we can see the distribution of the magnetization precession intensity, while in (c) the static demagnetizing fields, Eq.~(\ref{Eq:mod_hd}), are shown for the selected $\phi$ values. The number insets in the plots refer to the eigenmode visualizations.
\label{Fig:modes_demag}}
\end{figure*}

The simulation model has 65~nm at the thickest point, the width between the edges is 210~nm, and their rounding radius is 5~nm. 
Rounded edges of the waveguide are closely related to maintaining the integrity of the simulation by avoiding too thin elements. In addition, as shown in Ref.~\onlinecite{May2019RealisationLattice}, it is experimentally justified, and its simulated edge modes retain their physical properties.
 We used the following parameters of permalloy (Py): the saturation magnetization $M_\mathrm{S}=800$~kA/m, the exchange constant $A_\mathrm{ex}=13$~pJ/m, and the gyromagnetic ratio $|\gamma|=176$~GHz$\cdot$rad/T. To saturate the sample at all analyzed angles, we use an external magnetic field of $\mu_0\textbf{H}_\text{ext} = 3$~T.

\section{Micromagnetic simulations}
\label{Sec:mic_sim}
To comprehensively analyze the properties of the SW modes in the CS nanorod, we performed a series of numerical simulations in the Comsol Multiphysics software. It uses the finite element method (FEM) to solve coupled systems of partial differential equations, including the Landau-Lifshitz equation and Maxwell equations in the magnetostatic approximation.

All magnetic moments in numerically defined unit cells are modeled in the simulations as normalized unit vectors $\textbf{m}=\textbf{M}/M_\text{S},$ where $\textbf{M}$ is the spatio-temporal distribution function of the total magnetization. Then, neglecting damping as a parameter irrelevant to our analysis, the Landau-Lifshitz equation takes the form:
\begin{equation}
\frac{d\textbf{m}}{dt}=-\gamma\left[\textbf{m}\times\textbf{B}_\text{eff}\right],
\label{Eq:LLE}
\end{equation}
where $d\textbf{m}/dt$ is the time evolution of the reduced magnetization. The effective magnetic flux density field $\textbf{B}_\mathrm{eff}$ determines the direction around which the magnetization precesses and contains many system-related magnetic components. In our simulations we have assumed only the influence of the external magnetic field, exchange interactions and demagnetization.

The demagnetizing field $\textbf{H}_\text{d}$, contributes to the shape anisotropy in ferromagnets and to the SW dynamics. Since it is governed by Amp{\'e}re's law ($\nabla\times\textbf{H}_\text{d}=0$), the demagnetizing field can be derived from a gradient of the magnetic scalar potential $U_\text{m}$:
\begin{equation}
\textbf{H}_\text{d}=-\nabla U_\text{m},
\label{h_demag}
\end{equation}
\noindent which, inside the magnetic body, yields:
\begin{equation}
\nabla^2 U_\text{m}=\nabla\cdot\textbf{M}.
\label{Eq:nab2}
\end{equation}

All presented equations have been implemented in Comsol to solve the eigenproblem derived from Eqs.~(\ref{Eq:LLE})-(\ref{Eq:nab2}), assuming full magnetization saturation by the magnetic field, linear approximation, and analyzing only the CS planes. Assuming that ferromagnetic materials are saturated along the $i$-axis (orientation of $\textbf{H}_\text{ext}$), a linear approximation can be used to split the magnetization vector into static and dynamic (time $t$ and position $\textbf{r}$ dependent) components $\textbf{m}(\textbf{r},t) = m_i \hat{i} + \delta \textbf{m}(\textbf{r},t)\;\forall\;(\delta \textbf{m}\perp\hat{i})$, neglecting all nonlinear terms in the dynamic magnetization $\delta\textbf{m}(\textbf{r},t)$. For further methodological details, see Refs.~\cite{Mruczkiewicz2013StandingCrystals,Rychy2018SpinRegime}. Therefore, the numerical simulations were performed in two spatial dimensions with a triangular discretization of nearly 10000 cells. To visualize the static demagnetizing field on a two-dimensional $xy$-area in the form of a color map, we use the formula for its module:
\begin{equation}
{H}_\text{d}(x,y)=\sqrt{(dU_\text{m}/dx)^2+(dU_\text{m}/dy)^2}.
\label{Eq:mod_hd}
\end{equation}

To elucidate the SW spectra and SW dispersion relation in unsaturated CS nanorods, we use the finite difference method based micromagnetic simulation package -- Mumax3~\cite{Vansteenkiste2014TheMuMax3}. Here we solve the Landau-Lifshitz equation with the damping term (assuming damping coefficient $\alpha=0.0001$), leaving other $\textbf{B}_\mathrm{eff}$ terms the same as in Comsol simulations. To calculate the ferromagnetic resonance intensity spectra (Fig.~\ref{Fig:comp}) and dispersion relations (Fig.~\ref{Fig:disp}), we applied the Fast Fourier Transform (FFT). By applying the FFT, we were able to convert the time and space domain signals from our simulations into frequency and wave vector domain spectra and determine the resonant frequencies and SW mode profiles of CS waveguides.

Mumax3 field-rotation simulations were discretized by $256 \times 128 \times 1$ cells, each $0.92 \times 0.90 \times 1$ nm$^3$ in size, along the $x$, $y$, and $z$ axes, respectively. Periodic boundary conditions (PBC) were applied along the $z$-axis to mimic an infinitely long system. To excite the SW dynamics, we used a homogeneous in-space microwave magnetic field $\textbf{h}(z,t)=[h_0,h_0,h_0]\sinc{(2\pi f_\text{cut}t)}$ with amplitude $h_0=0.015H_{\text{ext}}$ and $f_\text{cut}= 100$ GHz.

For the dispersion relation simulation, however, the computational volume without PBC had to be increased to 20~\textmu m, which required the discretization reduction to $64 \times 64 \times 5120$ cells. To avoid reflections of SWs at the waveguide ends, an absorbing boundary condition was assumed. These adjustments did not affect the mode profiles, although a minimal shift in frequencies was observed. Importantly, in none of the simulations was the unit cell larger than the exchange length, which is 5.69 nm for Py. The SWs were excited from the central part of the waveguide by applying the time- and space-dependent dynamic component of the magnetic field $\textbf{h}$, defined as:
\begin{equation}
\textbf{h}(z,t)=[h_0,h_0,h_0]\sinc{(2\pi k_\text{cut}z)}\sinc{(2\pi f_\text{cut}t)},
\label{Eq:sinc}
\end{equation}
where $h_0=0.015H_{\text{ext}}$.
By using Eq.~\ref{Eq:sinc}, we can apply broadband SW excitation in ranges of frequencies $f\in[0, f_\text{cut}]$ and wavevectors along the $z$-axis $k_z \in [-k_\text{cut}, k_\text{cut}]$. 

\section{Results \label{Sec:Results}}

\subsection{Dependence of SW spectra on the orientation of the magnetic field \label{Sec:Angle_dependence}}

\begin{figure}[htp]
\includegraphics[width=\linewidth]{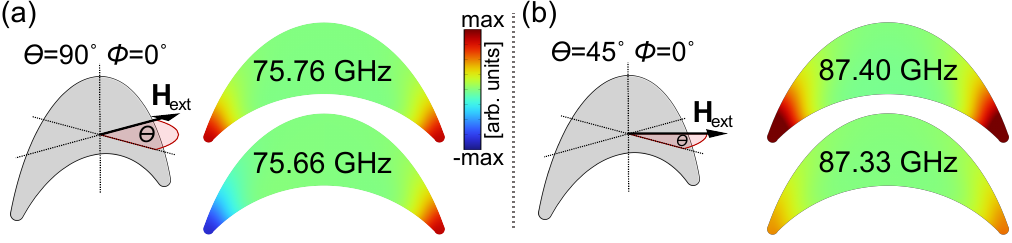}
\caption{Edge mode amplitude distribution for two first eigenfrequencies and external magnetic field polar angle equal to (a) 90$^\circ$ and (b) 45$^\circ$. In both cases the azimuthal angle remains 0$^\circ$, and the colors represent the sum of the dynamic components of the magnetization.
\label{Fig:symm_modes}}
\end{figure}

\begin{figure}[htp]
\includegraphics[width=\linewidth]{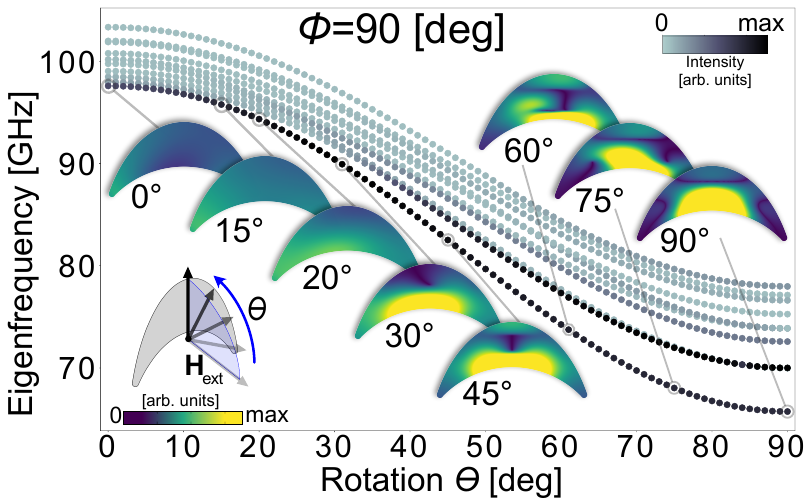}
\caption{Frequencies of the SWs in the CS nanowire as a function of the polar angle $\theta$ for a constant azimuthal angle $\phi=90^\circ$ of the 3~T external magnetic field (auxiliary diagram in the lower left corner). The dark color of the dots represents the resulting intensity of the eigenmodes calculated according to Eq.~(\ref{Eq:int}). The images of the cross sections show the distribution of the magnetization precession intensity for successively marked values of $\theta$.
\label{Fig:freq_theta}}
\end{figure}

We analyze the SW eigenmodes of the system shown in Fig.~\ref{Fig:geom}, saturated by the external magnetic field (3~T) directed at different angles $\phi$ and $\theta$. The results are shown in Figs.~\ref{Fig:modes_demag}-\ref{Fig:freq_theta}. They reveal interesting changes in the frequencies and distribution of the SW amplitude for different field configurations.

The analysis of changing the azimuth angle $\phi$ (at $\theta=90^\circ$) on the SW eigenmodes in Fig.~\ref{Fig:modes_demag} shows the evident influence of the system symmetry on their properties. This is manifested by the edge localization of some modes and the bulk concentration of others. The frequency shift with increasing $\phi$ is non-monotonic, revealing more and less favorable configurations for some applications. At $\phi=0^\circ$, the two low-frequency modes (see mode 1 in Fig.~\ref{Fig:modes_demag}~(a, b)) are the edge-localized SWs with antisymmetric and symmetric oscillations at the opposite edges of the nanorod, see Fig.~\ref{Fig:symm_modes}(a). Their frequency difference is only 100~MHz, indicating a weak coupling between the SW oscillations at the opposite edges. Interestingly, their responses to field rotation vary significantly, as the frequency of the symmetry branch increases linearly with increasing angle, from 75.76~GHz at $0^\circ$ to ${\sim}88$~GHz at $30^\circ$. The antisymmetric branch shows a completely different trend. We observe the transition from edge mode at $\phi=0^\circ$ (No.~1), to asymmetric single-edge localization at $\phi=45^\circ$ (No.~3), to low-frequency volume mode at $\phi=90^\circ$ (No.~5). The edge localization and its changes with magnetic field rotation can be explained by the enhancement of the demagnetizing field, which locally reduces the internal magnetic field, as shown in Fig.~\ref{Fig:modes_demag}~(c). Thus, the observed transition is strongly associated with the breaking of the symmetry of the internal magnetic field and the associated changes in the demagnetizing field.
For the examples in Fig.~\ref{Fig:modes_demag} labeled 3 and 4, we see that the demagnetizing field is strongly localized at the right edge; therefore, mode No. 3 has a lower frequency. On the other hand, modes 5 and 6 are volumetric, but both the magnetization and the demagnetizing field are not uniformly distributed over the cross section, which results in a lower frequency of mode 5 than 6, since the peak of the magnetization intensity coincides with the rise of the demagnetizing field.
Thus, the minima in the band structure for the polar angle $\phi=45^\circ$ are also related to the localization of these low frequency modes in one half of the CS, e.g. see mode No. 3 in Fig.~\ref{Fig:modes_demag}~(b). The demagnetizing field in the CS nanorod fluctuates with changing $\phi$, so does the frequency of the modes. Thus, the graph presented in Fig.~\ref{Fig:modes_demag}~(a) shows some crossing and anticrossing between different modes, but we have left their origin and interpretation for further study. Instead, in this paper we focus on analyzing the effect of field rotation on the CS system eigenfrequencies and SW propagation in them.

The above analysis of the external magnetic field rotation reveals the decoupling of the edge-localized SW modes. Consequently, the two edges can behave as separate, weakly dipolar coupled paths for SW guiding along the nanorod, with their frequencies controlled by the magnitude and orientation of the field. Furthermore, after the field rotation breaks the symmetry, CS nanorods can be used as two-channel SW waveguides capable of simultaneously supporting different frequencies on both sides.

Qualitatively, the situation is very similar for polar angles $\theta$ other than $90^\circ$, while the eigenfrequency values change significantly. In particular, the frequency of the edge mode increases from ${\sim}76$ to ${\sim}88$~GHz with a rotating magnetic field from $\theta=90$ to $45^\circ$ (Fig.~\ref{Fig:symm_modes}), since decreasing the demagnetizing field results in increasing the internal field. By changing $\theta$ from 90 to $45^\circ$ (and keeping $\phi=0^\circ$), we drive the first edge eigenmode to change its nature from antisymmetric to symmetric.

For this reason, we also decided to study the influence of the polar angle $\theta$ on the response of the ferromagnetic system. In this case, the symmetry between the geometry and the external magnetic field is maintained, i.e. for the constant angle $\phi=90^\circ$.

Fig.~\ref{Fig:freq_theta} shows the distinct and monotonic frequency drop (by ${\sim}30$~GHz for the lowest mode) along with the "skew" of the external magnetic field from the long axis of the nanowire. As for the variation of $\phi$, one can also observe a transition from the edge mode for small $\theta$ values to the volume mode already formed at about $\theta=20^\circ$.
The frequency drop is also accompanied by a greater separation of eigenmodes, especially the low-frequency ones, compared to the others. This is related to the increasing propensity of the system to generate an intense fundamental mode as $\theta$ grows. Furthermore, homogeneously precessing magnetization vectors are energetically more favorable for the field directed along a volume section of finite thickness. It will therefore oscillate with greater intensity and lower frequency.

In addition, the data markers on the frequency plots Fig.~\ref{Fig:modes_demag}~(a) and Fig.~\ref{Fig:freq_theta} were colored according to values of the following formula:
\begin{equation}
I=\left(\frac{1}{S} \int_{S}\delta\textbf{m}(\textbf{r},t)dS\right)^2
\label{Eq:int}
\end{equation}
where $S$ is an area of the nanorod's CS cross section. It defines the intensity of the eigenmodes, formulated to estimate their visibility in experiments, e.g. ferromagnetic resonance measurements. The most intense lines are associated with the fundamental mode (no phase change in the nanorod cross section); see mode 2 in Fig.~\ref{Fig:modes_demag}~(a, b). This is predictable since symmetric edge modes are enhanced by a strong demagnetization field, while volume modes occur over a larger area. However, azimuthal rotation of the field reduces the intensity and is only restored at $\phi>70^\circ$. Here we have two modes of comparable intensity at 65 and 70~GHz (see modes 5 and 6 in Fig.~\ref{Fig:modes_demag}~(b)). These modes have amplitudes concentrated in different parts of the inner and central parts of the nanorod with a nodal line perpendicular to $\textbf{H}_\text{ext}$. Their frequency splitting is due to different curvatures of the lower and upper nanorod edges and thus different demagnetizing fields (see Fig.~\ref{Fig:modes_demag}~(c) at $\phi=90^\circ$). Interestingly, rotating the field along the long axis (from $\theta=90^\circ$ to 0$^\circ$) decreases the intensity as they transform to the edge type mode (see Fig.~\ref{Fig:freq_theta}).

\subsection{Crescent-ellipse shape transformation}
\label{Sec:Ellipse}

\begin{figure}[htp]
\includegraphics[width=\linewidth]{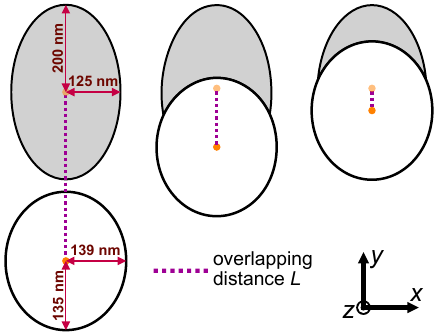}
\caption{Scheme of the analyzed shape transformation -- from ellipse to the crescent.
\label{Fig:scheme}}
\end{figure}

We further investigated the shape and curvature dependence of the analyzed cross section on the formation of low-frequency modes. Therefore, we performed a series of simulations for different geometries resulting from the superposition of two elliptical shapes, from a full ellipse to the crescent shape, controlled by a single parameter $L$. The scheme of the reasoning is shown in Fig.~\ref{Fig:scheme}. This allows us to capture the transition of the localization of the modes. In addition, we could also observe the effect of changing the geometry on the demagnetizing field. This analysis was performed for the 3~T external magnetic field along the $x$-axis.

\begin{figure*}[t]
\includegraphics[width=\linewidth]{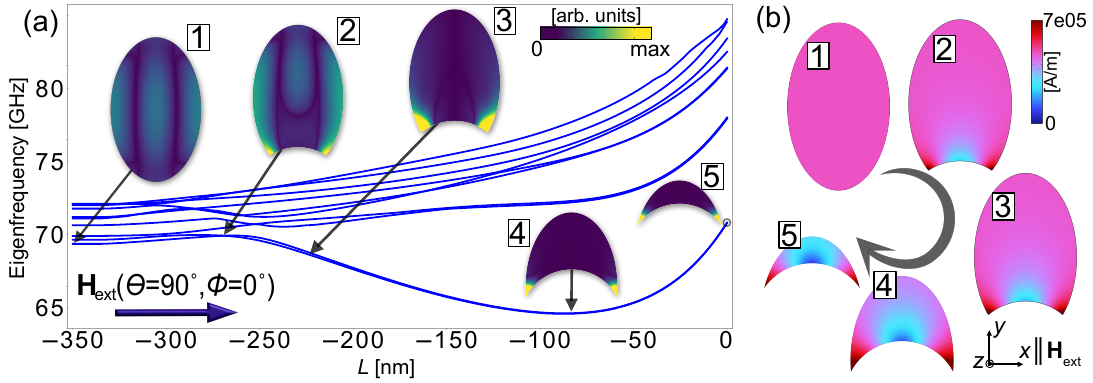}
\caption{Eigenanalysis of a variably shaped nanowire going from a full ellipse to the CS cross section (according to Fig.~\Ref{Fig:scheme}). In (a), the magnetization precession intensity distributions for the lowest eigenfrequency branch are shown for the five selected steps of overlapping distance $|L|=$~350, 270, 240, 90 and 0 nm. In (b) the distributions of the static demagnetizing field (see Eq.~(\ref{Eq:mod_hd})) are presented.
\label{Fig:trans}}
\end{figure*}

In Fig.~\ref{Fig:trans} it can be seen that the crossings of the individual frequency branches occur at the moment of transition, at $L\approx-270$~nm. Notable is also the non-monotonic dependence of the lowest frequency branches. By choosing the appropriate shape of the cross section, we can significantly influence their energy without changing the character of the modes (in this case, the edge mode excitation at the lowest frequency, i.e. 64.81~GHz, occurs for $L\approx-90$~nm). For $L>-90$ nm, the frequency of all modes increases with increasing $L$, indicating the dominant role of exchange interactions. An interesting phenomenon was also found for the first three pairs of eigenmodes -- we see that as the cross section is brought closer to the crescent shape, these modes degenerate. This is opposite to the behavior of Fig.~\ref{Fig:modes_demag}, where symmetric and antisymmetric edge modes split. It is also interesting from the point of view of designing such structures, since the geometry can tune the coupling between modes localized on opposite sides.

\subsection{Edge sharpness impact} \label{Sec:Edge}

Comparing the simulations for $\theta=90^\circ$ and $\phi=0^\circ$, we see that the first edge and volume modes (numbered 1 and 2 in Fig.~\Ref{Fig:modes_demag}) appear at 75.66 and 84.39~GHz, respectively. In the sharp-edged case (analogous to $L=0$~nm), the first eigenmode appears already at 70.88~GHz and is strictly localized at the edges. It shows a strong edge geometry/sharpness dependence of the eigenfrequencies obtained. This property provides another degree of freedom in this system to tune a magnonic spectrum. However, unlike the thickness, curvature and length of the nanowire, the shape of the edges is difficult to control experimentally. Therefore, in this analysis we only examine what differences can be expected and how essential this element is for the results of micromagnetic simulations. From experimental studies, it can be observed that the evaporation of a material on a polymer framework with an elliptical cross section~\cite{May2019RealisationLattice} leads to the formation of CS nanowires with partially rounded edges. This is mainly due to the roughness of the underlying resist of about 5~nm.

\begin{figure}[htp]
\includegraphics[width=\linewidth]{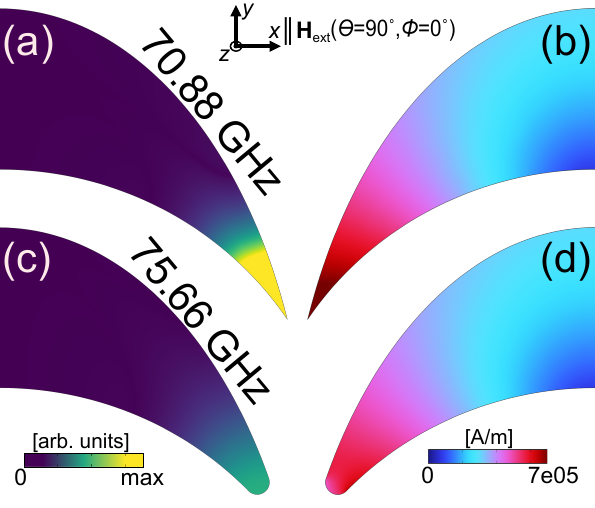}
\caption{Comparison of the localization of the SW modes and their frequencies (a, c) depending on the different edge sharpness and comparison with the corresponding demagnetizing fields (b, d). The color distributions for the modes represent the magnetization precession intensity from Eq.~(\ref{Eq:int}), while the demagnetizing field is from Eq.~(\ref{Eq:mod_hd}).
\label{Fig:edge_comparison}}
\end{figure}

In Fig.~\Ref{Fig:edge_comparison} it is evident that for the illustrative example ($\phi=0^\circ$, $\theta=90^\circ$), where the first eigenmodes are edge localized, there are obvious differences in their distribution and oscillation frequency, up to 4.78~GHz. As shown in the previous section, a strong demagnetizing field in thin regions, relative to a normally oriented magnetic field, causes the modes to be localized there. The crucial observation from Fig.~\Ref{Fig:edge_comparison} is that the frequency decreases significantly for the sharper edge case. This most likely means that the localization is still forced by the demagnetizing field (magnetostatic effect), since the exchange contribution would increase the frequency with persistent phase or amplitude inhomogeneities in the edge region.

\subsection{Unsaturated system} \label{Sec:Unsaturated}

\begin{figure}[htp]
\includegraphics[width=\linewidth]{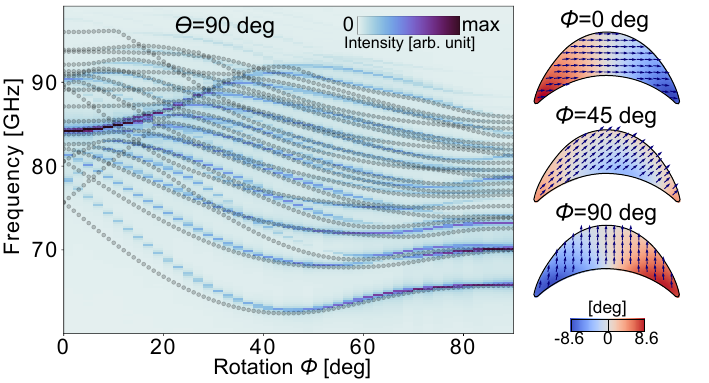}
\caption{Comparison of the CS nanowire's eigenanalysis results from Comsol (gray dots) and time-domain simulation results from Mumax3 (color scale) as a function of the azimuthal angle $\phi$ for the polar angle $\theta=90^\circ$ at external magnetic field 3~T. On the right, the static magnetization plots of the CS cross section from Mumax3 for three magnetic field configurations are shown. The color map visualizes the angle deviation of the static magnetization vector (arrows) from the direction of the external magnetic field.
\label{Fig:comp}}
\end{figure}

The eigenproblem simulations in Comsol assume a uniform static magnetization and focus on the change of field direction and its influence on the localization of harmonic SW modes and their frequencies. For this reason, a large 3~T magnetic field was assumed. An interesting observation is made when comparing the results of the frequency domain simulation with the results obtained in Mumax3 with the relaxation of the static magnetization distribution. As shown in Fig.~\ref{Fig:comp} (right panels), even at 3~T field, the infinite CS nanowire still maintains a non-uniform magnetization (up to 8.6$^\circ$ from the magnetic field orientation at the edges of the CS nanowire for $\phi=0$ and $90^\circ$), the strength and distribution of which vary with the magnetic field orientation. This leads to slight frequency differences (up to 2.5\%) between the Mumax3 and Comsol calculations (for low frequency modes at $\phi=0^\circ$ and higher frequency at $\phi=90^\circ$), while maintaining their qualitative agreement, as seen in Fig.~\ref{Fig:comp} (left panel). The intensity of a mode is also correlated for both solvers, which can be seen by comparing the intensity of the color map in Fig.~\ref{Fig:comp} with the intensity of the dots in Fig.~\ref{Fig:modes_demag}~(a). The unexpectedly large saturation field drew our attention to the use of CS waveguides for the propagation of high-frequency SWs, whose frequency will be strongly tunable with the propagation direction (chiral anisotropy) and the value and direction of the external field.

\subsection{Dispersion relation} \label{Sec:Dispersion}

Curvilinear magnetism, and in particular the propagation of SWs in cylindrical nanotubes, is the subject of analysis in recent studies~\cite{Otalora2016Curvature-InducedDispersion,Otalora2017AsymmetricCurvature,Streubel2016MagnetismGeometries,Korber2022CurvilinearStudy}. They focus on the influence of the magnetization chirality, forced by the geometry, on the dispersion relation of SWs, obtaining different frequency values with the same wavenumber but propagating in opposite directions. By analogy with the cylindrical cross section of a nanotube, we can assume that the magnetization in the CS nanorods spreads along their curvature at low magnetic fields, giving rise to SWs with chiral properties.
\begin{figure}[htp]
\includegraphics[width=\linewidth]{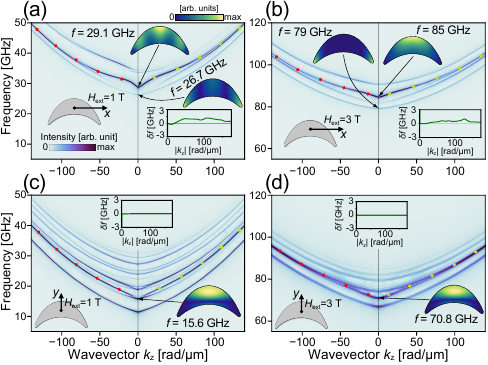}
\caption{Dispersion relations of CS nanorods for (a, c) 1~T and (b, d) 3~T external magnetic field directed along the $x$-axis (a, b) and the $y$-axis (c, d) for 1~T and (d) for 3~T. The wave vector is directed along the $z$-axis. Nonreciprocity is shown in insets as the frequency differences $\delta f$ between the most intense branches for positive (yellow dots) and negative (red dots) wavevectors. The plots also include visualizations of volumetric and edge modes (where they occur) for $k_z=0$.
\label{Fig:disp}}
\end{figure}
In Fig.~\ref{Fig:disp} we show the dispersion relations for a wavevector directed along the $z$-axis for two values of the external magnetic field, 1~T and 3~T, and its two orientations, $\phi=0$ and $90^\circ$ (for both $\theta=90^\circ$).

The results for the field directed along the $x$-axis (Fig.~\ref{Fig:disp}~(a, b)) show a clear, field-value dependent nonreciprocity for the bulk mode, represented by the inset plots $\delta f(|k_z|)=f(k_z) - f(-k_z)$. The highest value of this function for the bulk mode at 1~T along the $x$-axis is $\max(\delta f) \approx1.35$~GHz, and for the analogous case at 3~T -- $\max(\delta f) \approx1.03$~GHz. Interestingly, as we can see in Fig.~\ref{Fig:disp}~(a) and (b), the lowest frequency bands are the edge modes.
In the case of 3~T (Fig.~\ref{Fig:disp}~(b)), this mode is strictly edge localized (see also the mode No.~1 in Fig.~\ref{Fig:modes_demag}).  
Simulations also show that this mode has a symmetric parabolic dispersion relation, indicating the propagation nature of this excitation, which supports the thesis of a two-channel SW conductor in structures with CS cross sections. For smaller fields (e.g., 1~T in Fig.~\ref{Fig:disp}~(a)), the demagnetization exceeds the external magnetic field and the magnetization rotates tangentially to the edge of the nanowire, causing the amplitude to spread over the volume and lose its edge character. This results in a small nonreciprocity also for this mode. On the other hand, at the vertical magnetic field orientation, shown in Fig.~\ref{Fig:disp}~(c, d), there is a perfectly symmetric dispersion relation for both 1~T and 3~T field values.

The peculiar dependence of the SW dispersion relation on the value and direction of the magnetic field described in this section is a direct effect of the curvature when the magnetic field breaks the shape symmetry and its low amplitude does not allow it to saturate, leading to a quasi-chiral magnetization distribution. Therefore, the nonreciprocity can be explained by the fact that the static magnetization is distributed in a nonlinear way -- the smaller the field value, the more significant the nonreciprocity. Interestingly, we are still able to observe detectable nonreciprocity in the CS nanorods at fields as high as 3~T (asymmetrically directed). A novelty of the presented systems is the wide operating frequency range and the dynamic tunability using an external magnetic field. Another interesting step may be to test their operation in small fields (or even remanence), where the chirality and thus the nonreciprocity should be stronger and the SW frequencies lower. In addition, it is worth noting that the fabrication of structures with a CS cross section is less expensive than that of nanotubes, which is also an invaluable parameter for future applications.

\section{Conclusions}
In this study, an infinitely long waveguide with a CS cross section was investigated. The magnetic response of the system was examined (by eigenfrequency analysis) for different angles of the external magnetic field and the ratio of the edge curvature to the total volume of the cross section. From the obtained results we conclude that the dynamic manipulation of the field direction significantly changes the frequency and character of the eigenmodes, especially the low frequency ones, shifting from edge to volume localization. The critical factor in this transition is the effect of the symmetry breaking and the magnitude of the internal demagnetizing field in the edge and volume regions, respectively. An analogous transition from the low-frequency edge to the volume mode was observed by gradually changing the cross section from an elliptical to a narrow crescent. However, the relationship between shape and frequency remained non-monotonic, allowing the identification of the parameters analyzed to obtain the desired SW localization at a given frequency.

The analysis of CS structures is based on experimental research. Here, we numerically demonstrate that very long nanowires with such cross sections are interesting objects of study, allowing us to better understand magnonic effects in complex nanostructures. In 3D systems, where the nanorod elements are oriented at different angles to the external magnetic field, the magnetic effects resulting from the cross sections have a crucial impact on the global dynamic properties. There are also interesting aspects of CS waveguides whose magnetization is not fully saturated and is distributed along the curvature of the structure. For certain angles of incidence of the external magnetic field (in particular $\phi=0^\circ$ and $\theta=90^\circ$) it corresponds to a quasi-chiral-like texture and thus to an asymmetric dispersion relation for SWs propagating along its long axis. In addition, it was found that the saturation of this structure (in this configuration) is surprisingly high, which may favor the non-reciprocal propagation of high-frequency SWs.

An in-depth understanding of SW performance in waveguides with non-trivial shapes and cross sections is essential to understand the collective effects and advantages of using them in complex magnonic circuits. Ultimately, this research demonstrates the unique nature of CS-section nanorods and their real potential for use in future magnonic devices, where waveguides enabling advanced SW dynamics control will be critical.

\section*{Acknowledgments}
The research leading to these results was funded by the National Science Centre of Poland, Project No. UMO-2020/39/I/ST3/02413. HR, UM and MK acknowledge the financial support of the Ministry of Science and Higher Education in Poland, Grant No. MEiN/2022/DIR/3203. SL acknowledges support from the Leverhulme Trust (RPG-2021-139) and EPSRC (EP/R009147/1). AB gratefully acknowledges funding from Nano Mission, DST, India, under Grant No. DST/NM/TUE/QM-3/2019-1CSNB.

\section*{Author declarations}

\subsection*{Conflict of Interest}
The authors have no conflicts to disclose.

\subsection*{Author Contributions}
\noindent \textbf{Mateusz Gołębiewski:} Conceptualization (equal); Formal Analysis (equal); Methodology (equal); Project Administration (equal); Software (lead); Validation (equal); Visualization (lead); Writing -- Original Draft Preparation (lead); Writing -- Review and Editing (equal). \textbf{Hanna Reshetniak:} Conceptualization (equal); Formal Analysis (equal); Methodology (supporting); Software (equal), Visualization (equal); Writing -- Original Draft Preparation (supporting). \textbf{Uladzislau Makartsou:} Formal Analysis (supporting); Software (lead); Validation (supporting); Visualization (equal); Writing -- Original Draft Preparation (supporting). \textbf{Arjen van den Berg:} Methodology (equal), Software (equal). \textbf{Sam Ladak:} Conceptualization (equal); Methodology (equal); Project Administration (equal); Supervision (equal); Validation (equal); Writing -- Review and Editing (equal). \textbf{Anjan Barman:} Conceptualization (equal); Methodology (equal); Project Administration (equal); Supervision (equal); Validation (equal); Writing -- Review and Editing (equal). \textbf{Maciej Krawczyk:} Conceptualization (equal); Formal Analysis (equal); Funding Acquisition (lead); Methodology (equal); Project Administration (lead); Resources (lead); Supervision (lead); Validation (supporting); Writing -- Review and Editing (equal).

\section*{Data availability}
The data that support the findings of this study are available from the corresponding author upon reasonable request.

\bibliography{references}

\end{document}